\documentclass{article}

\usepackage{arxiv}

\usepackage[utf8]{inputenc} 
\usepackage[T1]{fontenc}    
\usepackage{hyperref}       
\usepackage{url}            
\usepackage{booktabs}       
\usepackage{amsfonts}       
\usepackage{nicefrac}       
\usepackage{microtype}      
\usepackage{lipsum}
\usepackage{graphicx}
\usepackage{todonotes}
\usepackage{multirow}
\usepackage{authblk}
\graphicspath{ {./images/} }

\title{
Can questions summarize a corpus? \\ \large Using question generation for characterizing COVID-19 research
}

\author[1]{Gabriela Surita}
\author[1,2,3]{Rodrigo Nogueira}
\author[1,2]{Roberto Lotufo}

\affil[1]{\footnotesize School of Electrical and Computing Engineering, UNICAMP}
\affil[2]{\footnotesize NeuralMind Inteligência Artificial}
\affil[3]{\footnotesize David R. Cheriton School of Computer Science, University of Waterloo}

\begin{document}
\maketitle
\begin{abstract}

What are the latent questions on some textual data? In this work, we investigate using question generation models for exploring a collection of documents. Our method, dubbed \texttt{corpus2question}, consists of applying a pre-trained question generation model over a corpus and aggregating the resulting questions by frequency and time. This technique is an alternative to methods such as topic modelling and word cloud for summarizing large amounts of textual data.

Results show that applying \texttt{corpus2question} on a corpus of scientific articles related to COVID-19 yields relevant questions about the topic. The most frequent questions are ``\textit{what is covid 19}'' and ``\textit{what is the treatment for covid}''. Among the 1000 most frequent questions are ``\textit{what is the threshold for herd immunity}'' and ``\textit{what is the role of ace2 in viral entry}''. We show that the proposed method generated similar questions for 13 of the 27 expert-made questions from the CovidQA question answering dataset.

The code to reproduce our experiments and the generated questions are available at: \url{https://github.com/unicamp-dl/corpus2question}

\end{abstract}


\section{Introduction}
\label{sec:introduction}

Methods for exploring, summarizing, and automatically organizing large volumes of textual data are commonly used by information retrieval and natural language processing systems. They allow researchers to quickly grasp what are widely discussed topics on specific literature as well as capture directions for further exploration. Examples of such methods are word cloud~\cite{wordCloud}, topic modelling ~\cite{lda} and document clustering~\cite{clusteringBook, doc2vec}. Still, these techniques are known to have limitations. Word cloud and topic modelling rely purely on n-gram statistics of documents in the corpus, and thus their outputs are keywords isolated from their original contexts. Clustering based on document embeddings try to capture the whole document context into a condensed vector, but sometimes fail to differentiate between parts of a document. Also, clusters are represented by document-long texts, whose assimilation can be cognitively demanding for users that are not familiar with the topic. Hence, clustering methods are often used in conjunction with visualization or summarization techniques to make their outputs more concise to humans.

We propose a different method for exploring and compressing information of a corpus based on the questions its documents can answer. We expect questions to capture deeper context knowledge than n-grams, but yet be a diverse and natural description of a corpus. This paper tries to answer two core questions: can deep learning models generate relevant research questions on scholarly articles on a zero-shot setting (i.e., without any training data on that particular domain)? If so, are these questions suitable for a compact description of a collection of documents?

\begin{figure}

    \begin{minipage}{.48\textwidth}
        \centering
        {\Large\texttt{corpus2question} (ours)} \\~\\
        \begin{tabular}{clr}
            \toprule
            \textbf{\#} & \multicolumn{1}{l}{\textbf{question}} & \multicolumn{1}{c}{\textbf{count}} \\ 
            \midrule
            1  & what is the treatment for covid         & 3185 \\
            2  & what is covid 19                        & 2604 \\
            3  & what is covid-19                        & 1775 \\
            4  & what is sars cov                        & 1694 \\
            5  & what is the risk of covid               & 1479 \\
            6  & what is the impact of covid             & 1449 \\
            7  & what is covid                           & 1316 \\
            8  & what is the prevalence of covid         & 1084 \\
            9  & what is sars-cov                        & 991  \\
            10 & what is the mortality rate of covid     & 942  \\
            11 & what is the incubation period for covid & 924  \\
            12 & what is the sir model                   & 863  \\
            13 & what is sars-cov-2                      & 823  \\
            14 & what is the incidence of covid          & 805  \\
            15 & what is social distancing               & 683  \\ \bottomrule
        \end{tabular}
    \end{minipage}%
    \hspace{1mm}
    \begin{minipage}{.48\textwidth}
        \centering
        {\Large Topic Modelling - LDA} \\~\\
        \begin{tabular}{cp{0.9\textwidth}}
        \toprule
        \textbf{\#} & \textbf{topics}                                                     \\ \midrule
1                               & data model used using based                         \\
2                               & research social students new work                   \\
3                               & cov sars sars\_cov cells protein                    \\
4                               & patients treatment 19 covid covid\_19               \\
5                               & air 2020 temperature study data                     \\
6                               & cases number model 19 covid                         \\
7                               & study health participants 19 covid                  \\
8                               & social distancing public contact social\_distancing \\
9                               & covid 19 covid\_19 risk age                         \\
10                              & patients 19 covid covid\_19 disease                 \\
11                              & covid patients 19 covid\_19 patient                 \\
12                              & 19 covid covid\_19 sars cov                         \\
13                              & data model time results using                       \\
14                              & patients 19 covid covid\_19 study                   \\
15                              & economic countries food global crisis               \\ \bottomrule
        \end{tabular}
    \end{minipage}%
    \vspace{10mm}
    \begin{minipage}{1\textwidth}
        \centering
        {\Large Word Cloud} \\~\\
        \includegraphics[width=280px]{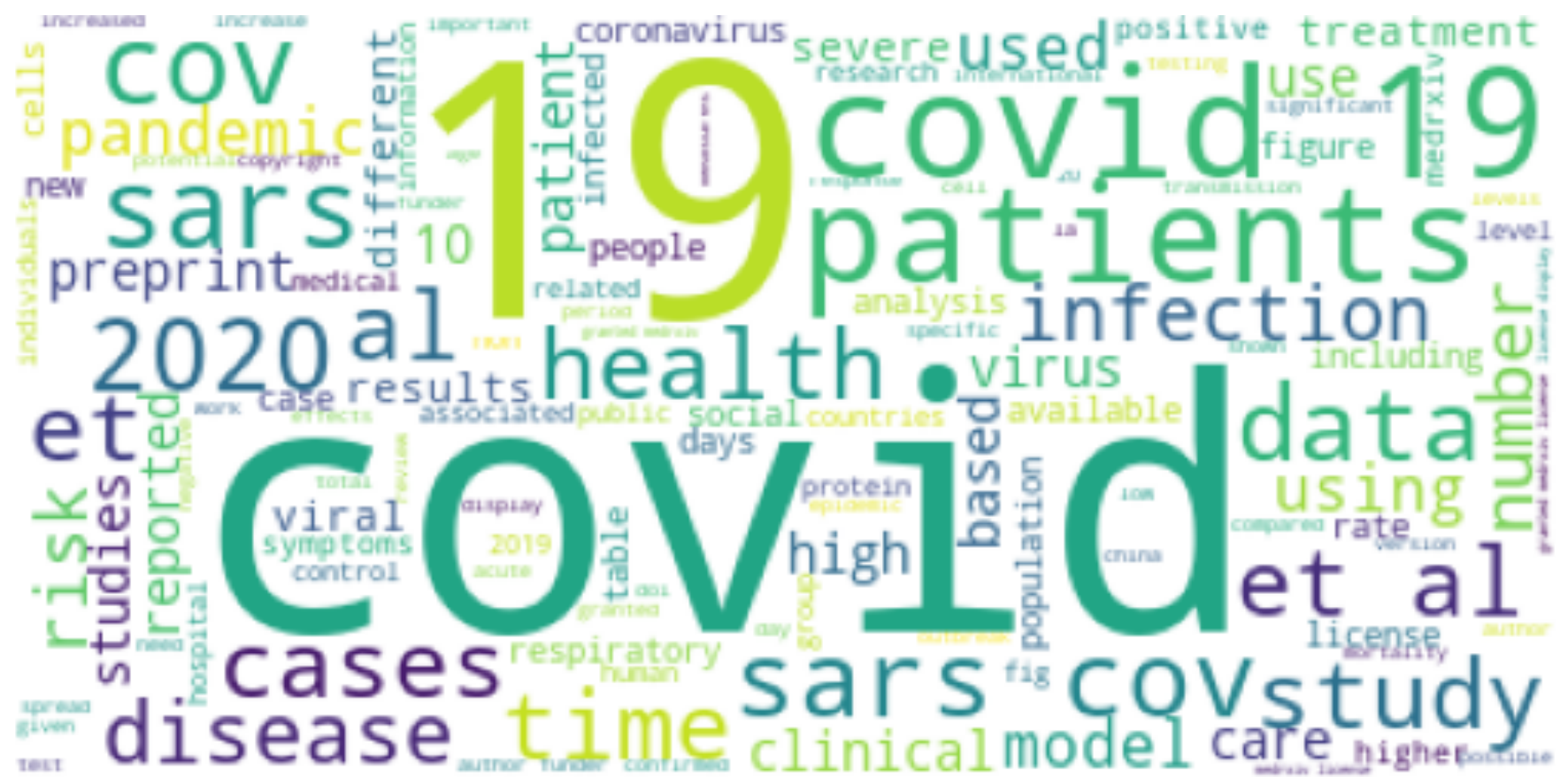}
    \end{minipage}%
    \vspace{5mm}
    \caption{Results for \texttt{corpus2question}, topic modelling using Latent Dirichlet Allocation (LDA) and word cloud over the CORD-19 dataset. The results for \texttt{corpus2question} show the 15 most frequent questions. The counts column represents the number of text spans that originated the question. Results for LDA uses 2-grams and 20 topics. We present 15 out of 20 topics with 5 most relevant n-grams for each topic.}
    \label{fig:main-results}
\end{figure}

We use existing question generation models, that, to our knowledge, have not been applied to out-domain data. To our knowledge there are no metrics to assess the quality of zero-shot question generation methods over large volumes of textual data, so we propose two quantitative methods to evaluate the model output in comparison to human-generated questions. We apply our method, called \texttt{corpus2question}, on CORD-19~\cite{cord19}, a collection of research papers related to COVID-19, and compare its questions against expert-generated questions from the CovidQA dataset~\cite{covidQA}. Finally, we include a qualitative discussion about the generations and some examples of how the data can be further aggregated. To demonstrate the relevance of the proposed method, we present its most frequent generated questions in Figure~\ref{fig:main-results}, alongside with results from alternative methods.

\begin{figure}
  \centering
    \includegraphics[width=380px]{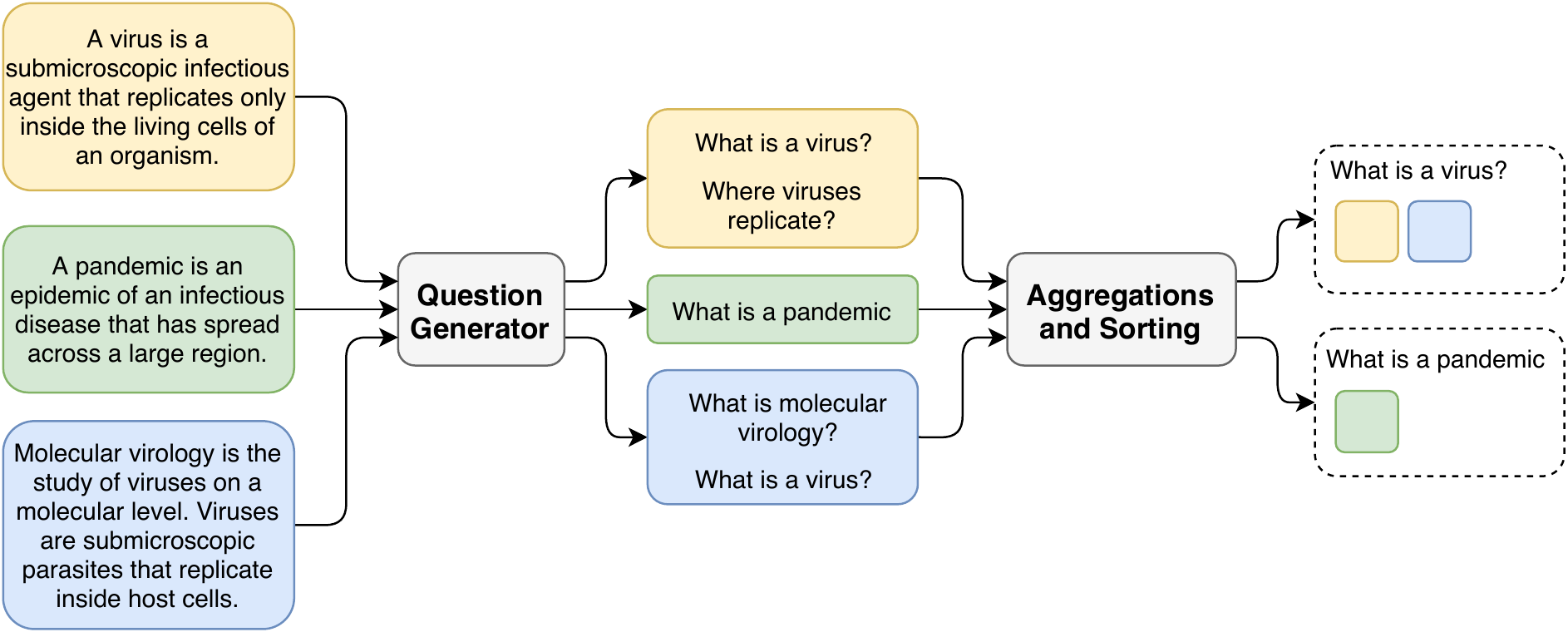}
  \caption{Illustration of the question generation pipeline followed by basic aggregations.}
  \label{fig:qg_pipeline}
\end{figure}

The rest of the paper is organized as follows: Section~\ref{sec:related-work} presents related work on methods for exploring datasets and question generation; Section~\ref{sec:methods} describes the models applied for question generation, the dataset preprocessing methods and two experiments proposed for evaluating the task; Section~\ref{sec:experiments} presents the results for the proposed experiments, as well as qualitative discussions about the generations; Section~\ref{sec:limitations} presents limitations of the current method and directions for future research. Finally, Section~\ref{sec:conclusions} concludes the work by reiterating key results.

\section{Related Work}\label{sec:related-work}

In this section, we present related work on dataset exploration techniques, more specifically, topic modelling and document clustering. We also describe the question generation techniques related to the proposed method. 

\subsection{Topic Modelling}

Topic modelling methods represent a corpus by underlying themes (described by n-grams) and map documents as combinations of these themes. One of the most relevant algorithms in this field is Latent Dirichlet Allocation (LDA)~\cite{lda}. LDA is a statistical generative method that given a number of topics and a tokenized text dataset produces a term distribution per topic and a topic distribution per document. 

LDA has been an inspiration for several variants that followed. Further work includes techniques for finding an optimal number of topics~\cite{ldaTopics} or improving the term definition~\cite{ldaTerms}. Other works try to combine LDA with word embeddings~\cite{lda2vec} or document embedding techniques~\cite{ldaEnsamble}.

Still, some publications point that LDA alone is sensitive to noise and can generate irrelevant topics on specific domains~\cite{ldaNoise}. Recent work has tried to address these issues with human-in-the-loop strategies~\cite{ldaCurated}. Others combine topic models with visualization tools like sankey charts to make generations more friendly to users~\cite{covidSee}.

\subsection{Document Clustering}

Document clustering is another widely used technique to represent large corpora. It consists of defining a vector representation for a document, a similarity metric between two vectors and then applying an unsupervised clustering algorithm~\cite{clusteringBook}. Known document representations are information-theoretical methods such as tf--idf~\cite{tf-idf} followed by a dimensionality reduction technique (such as PCA) and neural network methods such as doc2vec~\cite{doc2vec} and Sentence-BERT~\cite{sentenceBert}. 

These techniques are used in information retrieval systems and usually are suitable for search applications~\cite{bafna2016document}. Still, clusters are represented by document-long texts, which require additional inspection. Other work suggests that these techniques sometimes fail to replicate similarity as humans understand it~\cite{doc2vecEmpirical}. Some document representations also have low explainability, making it hard to know why a pair of documents is considered similar.

\subsection{Question Generation}

Extractive question answering is a widely studied task in NLP, with the SQuAD dataset~\cite{squad} being one of its most known examples. Given an input text and a question, the objective of this task is to answer the question using passages of the input text. The inverse task of extractive question answering is known as oriented question generation (QG)~\cite{qg}. Given an answer and a text passage, the objective is to generate a meaningful question. Question generation methods have been used in teaching assistant softwares~\cite{qgReview}, to improve academic writing~\cite{qgWriting} and to create synthetic datasets for improving question answering systems~\cite{qg4qa,alberti2019synthetic}.

As most text generation tasks, the state-of-the-art for this task is based on the transformer architecture \cite{transformer, qgTransformers}. Still, requiring an answer to generate a question limits the range of applications of such methods. Recent work suggests that transformers can generate meaningful questions given a text without any answer metadata \cite{qg-e2e}. A very similar task to question generation is known as query prediction~\cite{doc2query, doct5query}. In this task, given an input document, the model should generate a search engine query in which the document is relevant. In this setup, there is also no short answer span, or the whole document is considered as an answer. Search performance can then be improved by expanding the document with its predicted queries before indexing. 

\section{Methods}
\label{sec:methods}

In this section, we present the question generation method, as well as preprocessing and post-processing techniques and the proposed experiments to evaluate the model.

\subsection{Preprocessing and Post-processing}\label{sec:post-processing}

The first step of the preprocessing method consists of removing non-natural language passages using regex (e.g. email addresses, URLs, DOIs, citation tags and section numbers). We then apply sentence tokenization on every input document. We remove passages with only one sentence as they usually represent section names (i.e. Introduction). 

Before generation, we break documents into sentences and then group sentences into spans of 10 sentences with a stride of 5. For every span, we generate a single question using doc2query, which is described in Section~\ref{sec:methods-qg}.

After the generation, we apply a simple post-processing step that removes publisher names and questions containing the words ``\textit{preprint}'' and ``\textit{copyright}'' as these were frequent on the top questions.

\subsection{Question generation - doc2query}\label{sec:methods-qg}

The generation model consists of a pre-trained language model based on T5-base~\cite{t5} that has been fine-tuned on the MS MARCO passage ranking dataset~\cite{msmarco}. The dataset contains over 500k queries sampled from the Bing search engine. Queries are paired with at least one manually annotated relevant passage. The model has originally been used for document expansion through query generation~\cite{doct5query} and is available on Github. \footnote{https://github.com/castorini/docTTTTTquery} We do not apply any gradient updates to the publicly available model, i.e., we use it as an off-the-shelf model. For each input passage, we generate a single query using beam search decoding with 4 beams. 

\subsection{Per-document question accuracy}\label{sec:method-per-doc-acc}

\begin{figure}
  \centering
    \includegraphics[width=380px]{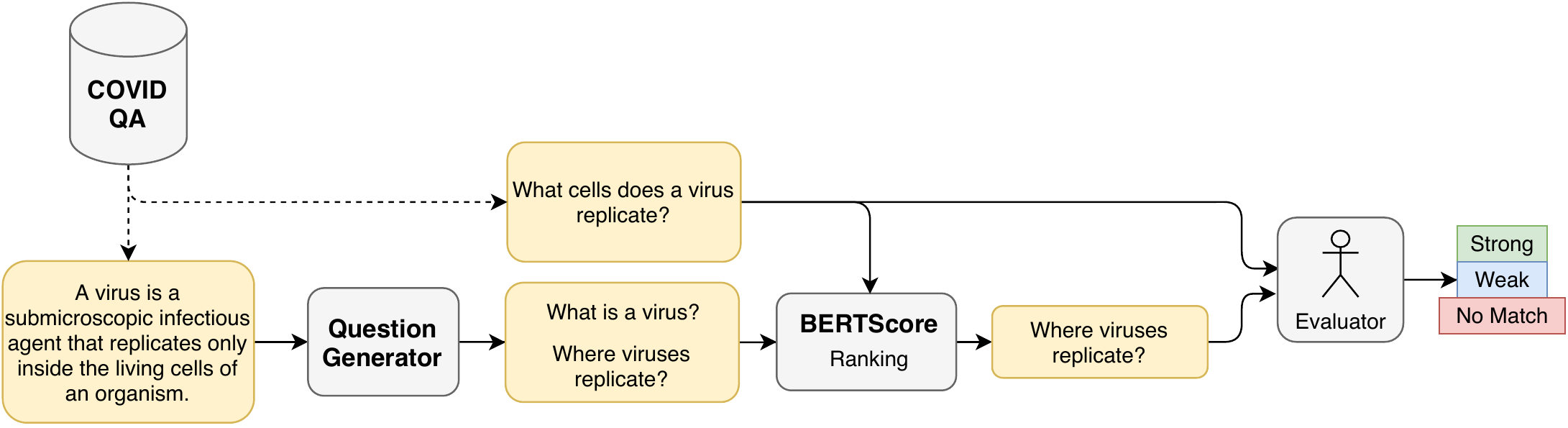}
    (a) 
    \\~\\
    \includegraphics[width=380px]{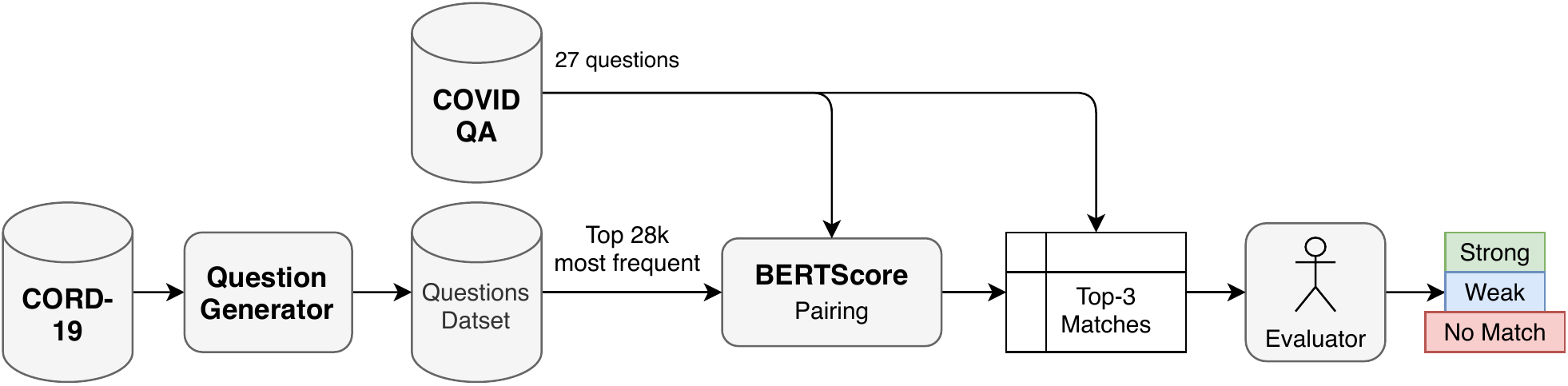}
    (b)
    \\
  \caption{Illustration of the per-document question accuracy (a) and the dataset question relevance (b) experiments. The first experiment captures the relevance of the question generated with respect to an expert reference for every document individually. The second experiment verifies the relevance of the most frequent generated questions with respect to a list of questions in a reference dataset.}
  \label{fig:covidqa-pipeline}
\end{figure}

Here we present a method for capturing the quality and relevance of the generated question for a single document regardless of other document generations. The data pipeline is presented in Figure~\ref{fig:covidqa-pipeline}(a). We use a dataset that has questions created by humans and their respective documents that contain an answer. For each document, we generate questions using the model described in Section \ref{sec:methods-qg}. Our goal is to validate for every document if at least one of the questions generated by the model is the same as the human-created question. We apply a ranking method based on BERTScore~\cite{bertScore} to extract the most similar questions to the reference question. It uses contextual embeddings from BERT~\cite{bert} to compute cosine similarity between candidate and reference questions. BERTScore has been shown to have closer correlation to human evaluations than other metrics such as BLEU and ROUGE. We present the reference and the top 3 ranked generated questions to a non-expert human evaluator that annotates the extracted questions as strong, weak, or no match with respect to the reference question.

BERTScore is used as a ranker so the evaluator does not have to search through all questions generated from the document, which makes the annotation process faster. Two levels of semantic matches are considered to capture the quality of the generation. A strong match implies that the question generated by the model is semantically the same question as the reference. A weak match is when the model generates a semantically broader question than the reference. To be considered a weak match, the response of the reference question should be contained on the generated question.

\begin{figure}[!htbp]

\fbox{\parbox{\textwidth}{

{\bf Reference:}
How does temperature affect the transmission of COVID-19?

\vspace{3mm}

{\bf Strong Match:}
What is the influence of temperature on the transmission and spread of COVID-19?

{\bf Weak Match:}
What is the influence of meteorological factors on the transmission and spread of COVID-19?

{\bf No Match:}
What is the influence of humidity on the transmission and spread of COVID-19?
}}

\caption{Examples of weak, strong, and no match evaluations. A strong match represents that the two sentences are semantically the same. A weak match represents that the generated question is broader than the reference, but within the same topic.}

\end{figure}

\subsection{Frequent question relevance}

This experiment tries to capture if important questions are frequently generated by the model. More specifically, if among the most frequent questions we can find questions from a human-made question dataset. The data pipeline is presented in Figure~\ref{fig:covidqa-pipeline}(b). We first generate questions for all documents of the dataset. Then we use BERTScore~\cite{bertScore} to measure how similar each question generated by the model is to a reference question. Finally, we present the reference and top 3 scored questions to a non-expert evaluator that annotates the generated question as strong, weak, or no match. 

Pairing is done again with a ranking function based on BERTScore to simplify the evaluation step, reducing human annotation workload to only 3 generated questions per reference question. The annotation labels are the same as the one described on the per-document experiment.

\section{Experiments and Discussion}
\label{sec:experiments}

Here we present the experiments to verify \texttt{corpus2question} as a viable technique for text exploration. In Sections \ref{sec:experiments-per-doc} and \ref{sec:experiments-relevance} we compare the questions generated by the model with questions produced by experts. In section \ref{sec:experiments-aggregations} we showcase some examples of frequency and time aggregations over our generated questions, and finally in Section \ref{sec:compare-lda} we compare the results produced by our model with LDA and word cloud results.

\subsection{Datasets}\label{sec:datasets}

We apply \texttt{corpus2question} on the CORD-19 dataset~\cite{cord19}, which is a collection of research papers on COVID-19 and related diseases. We used the release of August 26th, 2020, which contains 101,688 full documents. These papers are extracted from several sources and made available in text format. Documents are presented in JSON format, preserving structures such as paragraph breaks and section headers as separate passages.

We preprocess the dataset to remove articles about similar respiratory diseases that are not COVID-19. The motivation is to reduce the time spent on query generation, as well as to reduce noise on the dataset. For that we keep only articles with publication date after October 2019 and that include several covid-related terms such as COVID, SARS-CoV, SARS-2, Wuhan, and China. The resulting dataset contains 41,526 full papers.

For the quantitative experiments, we used CovidQA~\cite{covidQA} as our reference question dataset. CovidQA is bootstrapped from literature reviews made available on Kaggle over the CORD-19 dataset. Its version 0.2 contains 27 unique questions whose answers were found in 156 documents.

\subsection{Per-document question accuracy}\label{sec:experiments-per-doc}

On the per-document question accuracy experiment described in Section~\ref{sec:method-per-doc-acc}, we evaluated the 136 documents from the CovidQA that contains answers to 27 different questions. We found that 67 (47\%) of the best-ranked questions matched the reference questions, with 45 (33\%) strong and 22 (16\%) weak matches.

We also show that the accuracy drastically varies according to the question. Among the questions with the largest number of documents, reference question "Incubation period of the virus" found 22 strong matches out of 23. In contrast  "Sample size used in COVID-19 studies" found only 1 strong match among 26 documents. This result suggests that the model is more inclined to generate some annotated questions more than others. 

The lack of a baseline makes it hard to compare the relevance of this result, but the absolute number indicates that the model does capture at least half of the human-made generations.

\subsection{Frequent question relevance}\label{sec:experiments-relevance}

For the question relevance experiment, we first generated questions for all the selected examples of the CORD-19 dataset, with a total of 736,219 generations and 470,749 unique questions. Among those, we filtered those that showed up on at least 3 different documents, summing a total of 28,362 questions (3,8\%). We compare these to the 27 questions on the CovidQA dataset.

We found 13 (48 \%) question matches out of 27 after pairing. With 8 strong and 5 weak matches. We show that the model was again able to generate almost half of the annotated questions among its most frequent generations.

We also present in Table~\ref{tab:match-examples} some strong and no match examples. We see that while strong matches usually occur for short and simple questions, there are generally no matches for specific and targeted questions.

\begin{table}
\centering
\caption{Samples of strong and no matches extracted from the frequent question relevance experiment.}
\label{tab:match-examples}
\begin{tabular}{l l}
\toprule
\textbf{Reference}                                                & \multirow{2}{*}{\textbf{Tag}} \\
\textbf{Generation}                                               &                               \\ \midrule 
What is the incubation period of the virus?                       & \multirow{2}{*}{Strong}       \\ \vspace{2mm}
what is the incubation period of the virus                        &                               \\ 
What is the proportion of patients who were asymptomatic?         & \multirow{2}{*}{Strong}       \\ \vspace{2mm}
what is the proportion of asymptomatic cases                      &                               \\ 
Effects of temperature on the transmission of COVID-19	          & \multirow{2}{*}{Strong}       \\ \vspace{2mm}
what is the effect of temperature on the spread of covid          &                               \\ 
What is the RR for severe infection in COVID-19 patients with hypertension?            & \multirow{2}{*}{None}       \\ \vspace{2mm}
what is the incidence of thrombotic complications in patients with covid 19 infection? &          \\ 
What is the mortality rate for COVID-19 patients with hypertension?                    & \multirow{2}{*}{None}       \\ \vspace{2mm}
what is the mortality rate for covid patients                     &                               \\ 
What is the HR for death in COVID-19 patients with diabetes?      & \multirow{2}{*}{None}       \\ \vspace{2mm}
what is the leading cause of mortality in patients with covid 19? &                               \\
\bottomrule
\end{tabular}
\end{table}

\subsection{Question aggregation methods}\label{sec:experiments-aggregations}
\label{sec:experiment-showcase}

In this section, we present further explorations we made over the generated questions that have not been validated through quantitative experiments. We use this space to showcase some of the exploration possibilities enabled by the proposed technique with qualitative discussions about the quality of the generations.

\subsubsection{Most frequent questions}

In Figure~\ref{fig:main-results}, we show the top 15 most frequent questions generated from the corpus. We note that frequently generated questions are concise descriptions of the main theme of the corpus. Some examples are related to the disease name (COVID-19), the virus (SARS-CoV-2), treatments, the resulting pandemic, and also other related topics such as symptoms, the outbreak, and social distancing. Among the top 1000 most frequent questions, we show some interesting ones in Table~\ref{tab:top-1000}. We argue that those questions cannot be immediately perceived by someone less familiar with the corpus. Hence, \texttt{corpus2question} might help in generating hypothesis or developing underexplored areas of research.

\begin{table}[]
\centering
\caption{Samples from the 1000 most frequent questions from the CORD-19 dataset.}
\label{tab:top-1000}
\begin{tabular}{rlr}
\toprule
\multicolumn{1}{l}{\textbf{\#}} & \textbf{question}                                  & \multicolumn{1}{l}{\textbf{counts}} \\ \midrule
177 & what is the long term effect of covid          & 105 \\
246 & what is the cytokine storm in covid            & 84  \\
265 & what is the most common complication of covid  & 79  \\
322 & what organs are affected by sars               & 70  \\
338 & what is the threshold for herd immunity        & 67  \\
365 & what is the cause of death for covid           & 63  \\
413 & what is the most common comorbidity of covid   & 58  \\
431 & what is the incidence of covid in children     & 57  \\
524 & how long does it take to recover from covid    & 49  \\
537 & what is the basic reproduction number of covid & 47  \\
559 & what is the treatment protocol for covid       & 46  \\
567 & what is the clinical diagnosis of covid        & 45  \\
593 & what is the cutaneous manifestation of covid   & 44  \\
623 & what is the epitope of sars-cov-2              & 42  \\
671 & what is the role of t cells in sars            & 39  \\
692 & what is the role of ace2 in viral entry        & 39  \\
734 & What is the effect of temperature of the spread of covid & 37  \\
736 & what is the incidence of aki in covid          & 37  \\
748 & how long does it take for sars to show up in blood & 36 \\
757 & what is the sensitivity of a chest ct scan     & 36  \\
807 & what is the vertical transmission of sars      & 34  \\
898 & what is the innate immune response to sars     & 31  \\
907 & what is the mechanism of action of ace2        & 31  \\
972 & what is the role of ace2 in lung injury        & 29  \\ \bottomrule
\end{tabular}
\end{table}

\subsubsection{Questions over time}

\begin{figure}
  \centering
    \includegraphics[width=360px]{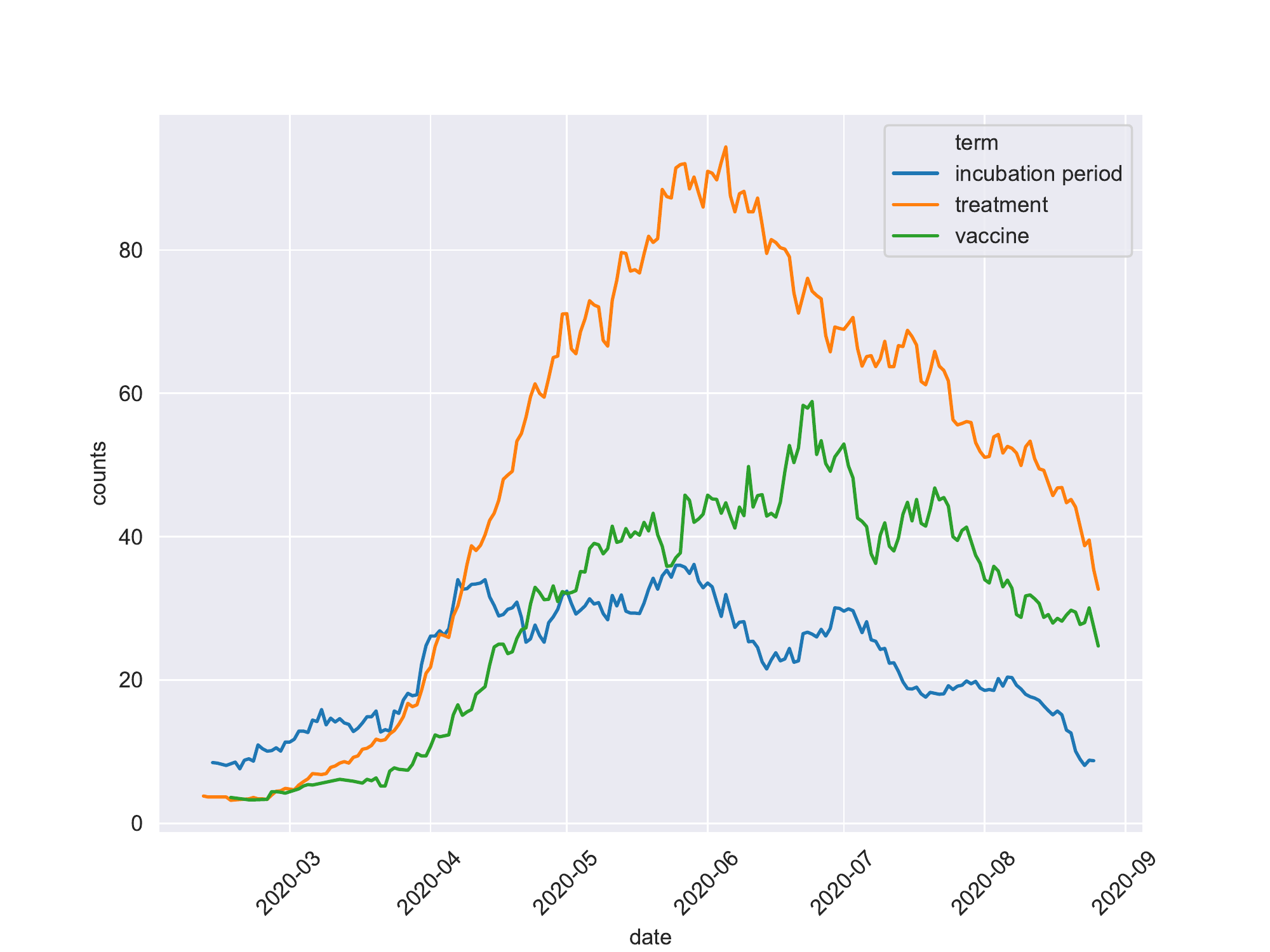}
  \caption{Number of questions related to incubation periods, treatments and vaccines for COVID-19. The horizontal axis denotes the publication date of the paper that the question was generated from.}
  \label{fig:treatements}
\end{figure}

Another interesting property is that we can plot the question frequency over time using the article publication date. Figure~\ref{fig:treatements} shows the number of questions related to incubation periods, treatments and vaccines over time. To select those questions we check if the question contain the words ``incubation period'', ``treatments'' or ``vaccine''. Since the beginning of the pandemic, the search for a treatment has been steadily increasing. The interest for incubation periods has appeared early on 2020, but has not grown substantially since then, suggesting that this topic reached consensus. We see that the interest for a vaccine is delayed when compared to the interest in incubation periods, but the former has surpassed the latter in the last couple of months. 

\subsection{Comparison with LDA and Word Clouds}\label{sec:compare-lda}

Here we compare our method with more traditional techniques such as LDA and word cloud. We present results for topic modelling over CORD-19 in Figure~\ref{fig:main-results} obtained with LDA using tri-grams and 20 topics, with the same pre and post-processing techniques described on Section~\ref{sec:post-processing}. We manually remove 5 non-sensical topics (e.g ``\textit{et al et\_al}'', ``\textit{la en el}'', and ``\textit{doi author funder}''), which results in a total of 15 topics being presented. We also present a word cloud using tri-grams.\footnote{\url{https://github.com/amueller/word_cloud}} LDA and word cloud techniques were able to identify a few buzzwords from the subject, such as ``covid'', ``sars'', ``health'', and ``pandemic'', but they fail to provide more context. For example, someone not familiar with the theme will probably find difficult to learn anything from the keywords ``data model time results using'' (topic 13 from Figure~\ref{fig:main-results}). However, if presented with the question ``\textit{what is the sir model}'', one could infer that the data refers to the SIR model for infectious diseases. We argue that this suggests that methods such as \texttt{corpus2question} that capture local context, can be useful text exploration tools.

\subsubsection{Questions associated with LDA}

We can use questions as an alternative representation of documents for methods like topic modelling and clustering. In Table~\ref{tab:questions-lda} we present the topics obtained by applying LDA to questions generated by \texttt{corpus2question}. In this experiment, every question is used as a document. We also present the most relevant question for every topic, selected by calculating the probability of the question belonging to the topic multiplied by the frequency of the question.

While we do not have a metric to compare if the topics generated by this method explain the dataset better than the ones computed over the original corpus, we can highlight some interesting properties of these results. First, we obtain a concise representation of a topic in the format of a question (e.g \textit{``rate covid ct mechanism rate\_covid''} can be represented by \textit{``what is the mortality rate of covid''}). Second, we substantially increase the variability of the questions in comparison to using only frequency aggregation presented in Figure~\ref{fig:main-results}, while still maintaining relevance within the topic.

\begin{table}
\centering
\caption{Topic modelling (LDA) applied to questions generated by \texttt{corpus2question}. We present 20 topics and their respective 5 most relevant n-grams. For every topic, we also present one relevant question selected using the frequency of the question among all generated questions multiplied by the probability of the question belonging to the topic.}
\label{tab:questions-lda}
\begin{tabular}{cll}
\toprule
\multicolumn{1}{l}{\textbf{\#}} & \textbf{topics}                            & \textbf{question}                           \\ \midrule
1                               & function pcr virus mers sars               & what is rt pcr                              \\
2                               & purpose covid prevalence cause sensitivity & what is the prevalence of covid             \\
3                               & treatment vaccine sars cov2 sars\_cov2     & what is the treatment for covid             \\
4                               & role health hcq care study                 & what is the clinical manifestation of covid \\
5                               & important role viral replication il        & what is il-6                                \\
6                               & normal range normal\_range sars wuhan      & what is the most common symptom of covid    \\
7                               & risk factor risk\_factor remdesivir icu    & what is the risk factor for covid           \\
8                               & covid age average use ppe                  & why do people wear masks                    \\
9                               & pandemic period impact sars incubation     & what is the incubation period for covid     \\
10                              & rate covid ct mechanism rate\_covid        & what is the mortality rate of covid         \\
11                              & does long long\_does coronavirus mortality & what is social distancing                   \\
12                              & covid 19 covid\_19 did pandemic            & what is covid 19                            \\
13                              & scale temperature rna relationship social  & what is the genotype of covid               \\
14                              & sars does long test long\_does             & how long does it take for sars to spread    \\
15                              & used protein spike number coronavirus      & what is the basic reproduction number       \\
16 & covid clinical epidemic pandemic important & what is the clinical course of covid \\
17                              & difference network vitamin target rbd      & what is vitamin d deficiency                \\
18                              & model purpose covid crisis disease         & what is the sir model                       \\
19                              & definition effect necessary air pollution  & what is the definition of covid             \\
20                              & sars cov sars\_cov role ace2               & what is sars cov                            \\ \bottomrule
\end{tabular}
\end{table}

\section{Limitations}\label{sec:limitations}

In this section, we highlight important limitations of the proposed method. 

\subsection{No definition of irrelevant or redundant questions}

In our experiments, we do not evaluate irrelevant, redundant or unrelated questions. One could argue that the probability of generating matching questions for a document increases with the number of generated questions just by chance. Still, the most frequent questions presented in our experiments suggest that meaningful questions are more frequent than noisy ones. 

To prevent matching from improving just by having more passages, we chose the largest span window that fits in our GPU and a single generation per span. The decision for not optimizing hyperparameters was to prevent improving over the proposed metrics while generating too many irrelevant or redundant questions. A  metric that accounts for "bad" questions would allow a better selection of hyperparameters such as window size, stride, and the number of generations per window.

\subsection{Long tail of infrequent questions}

From the 720,545 generated questions, 391,910 (54\%) have a frequency of 1, that is, they are generated from only one passage. This statistic shows that while frequent questions are suitable for direct aggregation there is still a large number of infrequent questions. We do not have a clear explanation of how this relates to the question generation model or the source documents. Using better aggregation techniques or highlighting representative questions of a group of infrequent questions may be an important future work direction.

\subsection{Influence of the fine-tuning dataset}

We do not know to what extend the dataset used to fine-tune the question generation model might have influenced the results as we use on all experiments an off-the-shelf model fine-tuned on the MS MARCO dataset. We considered using the same generation model but fine-tuned on the SQuAD 2.0 dataset~\cite{squad2}. Preliminary experiments showed that this model generated long and factual questions, not suitable for aggregation. To illustrate the effect, in Table~\ref{tab:dataset-examples}, we present some sample questions extracted from each dataset.

It seems that the quality of the fine-tuning dataset affects the results, but we do not investigate this further. We suspect that a task-specific dataset based on literature reviews of scholarly papers might be highly beneficial for this task.

\begin{table}[!htbp]
\centering
\caption{Questions from the SQuAD 2.0 and the MS MARCO datasets. Questions from MS MARCO are shorter and broader than the ones from SQuAD.}
\label{tab:dataset-examples}
\begin{tabular}{cl}
\multicolumn{1}{l}{\textbf{Dataset}} &
  \textbf{Examples} \\ \hline
SQuAD 2.0 &
  \begin{tabular}[c]{@{}l@{}}How did the black death make it to the Mediterranean and Europe?\\ What word is the word pharmacy taken from?\\ How would the word apothecary be viewed by contemporary English speakers?\end{tabular} \\ \hline
MS MARCO &
  \begin{tabular}[c]{@{}l@{}}what makes your hands burn\\ what is medicare summary notice\\ best anti inflamatory after surgery\end{tabular} \\ \hline
\end{tabular}
\end{table}

\subsection{Influence of the model size}

All experiments presented are done with a small model compared to state-of-the-art models for other tasks. Recent work has demonstrated that size matters on text generation tasks~\cite{t5, gpt3}. Further investigating the influence of models sizes and architectures may also be an important future step to improve the quality of generations.

\section{Conclusions}
\label{sec:conclusions}

We introduced \texttt{corpus2question}, a method that uses question generation models followed by frequency aggregations to summarize corpora. We demonstrate that this method applied to the CORD-19 dataset yields several relevant questions about COVID-19. We show that \texttt{corpus2question} is able to capture contextual information about documents when compared to other exploration methods such as word cloud and topic modelling. We also propose two quantitative experiments to compare the questions to human-generated questions that may be used to evaluate future research on the subject. Finally, we conclude our work by presenting its limitations and opportunities for future work in this field. Due to the technical nature of the CORD-19 dataset, we note that further evaluating the relevance and accuracy of frequent questions generated by the model may require assistance from experts with medical and biomedical backgrounds.

\bibliographystyle{unsrt} 
\bibliography{references} 

\end{document}